\documentclass[prr,twocolumn,superscriptaddress,longbibliography,nofootinbib]{revtex4-1}
\usepackage{latexsym}
\usepackage{lineno}
\linenumbers
\usepackage{hyperref}
\usepackage{url}
\usepackage[caption=false,font={normalsize}]{subfig}
\usepackage{graphicx}
\usepackage{verbatim}
\usepackage{multirow}
\usepackage{amsmath, bm}
\newcommand{\beq}{\begin{eqnarray}}
\newcommand{\eeq}{\end{eqnarray}}
\usepackage{mathrsfs}
\usepackage{float}
\usepackage{bbm}

\usepackage{mathtools}
\usepackage{slashed}
\usepackage{physics}	
\usepackage{graphicx}   
\usepackage{epstopdf}
\usepackage{hyperref}   
\usepackage{bbold}
\usepackage{wasysym}
\usepackage{amssymb}
\usepackage{soul}
\usepackage[symbol]{footmisc}

\usepackage{tikz}
\usetikzlibrary{decorations.pathmorphing}
\usetikzlibrary{shapes.misc}
\tikzset{cross/.style={cross out, draw=black, minimum size=8*(#1-\pgflinewidth), inner sep=0pt, outer sep=0pt},
cross/.default={1pt}}
\usetikzlibrary{patterns,math}
\newcommand{\RN}[1]{%
  \textup{\uppercase\expandafter{\romannumeral#1}}%
}

\nolinenumbers
\begin{document}
\title{Signatures of Green's function zeros and their topology using impurity spectroscopy
}
\author{Sayan Mitra}
\affiliation{Department of Physics and Astronomy, Iowa State University, Ames, Iowa 50011, USA}
\author{Fang Xie}
\affiliation{Department of Physics and Astronomy,
Extreme Quantum Materials Alliance,
Smalley-Curl Institute,
Rice University, Houston, Texas 77005, USA}
\author{Marek Kolmer}
\affiliation{Ames National Laboratory, U.S. Department of Energy, Ames, Iowa 50011, USA}
\author{Qimiao Si}
\affiliation{Department of Physics and Astronomy,
Extreme Quantum Materials Alliance,
Smalley-Curl Institute,
Rice University, Houston, Texas 77005, USA}
\author{Chandan Setty$^\dagger$}
\affiliation{Department of Physics and Astronomy, Iowa State University, Ames, Iowa 50011, USA}
\affiliation{Ames National Laboratory, U.S. Department of Energy, Ames, Iowa 50011, USA}

\begin{abstract} 
Topology without quasiparticles has emerged as a key framework for understanding Mott insulators, where Green's-function zeros encode nontrivial topological structure. Yet, experimental detection of these zeros represents a challenge. 
Using exact diagonalization of the one-dimensional Hubbard model with an impurity and Zeeman field, supported by exact analytic results, we show that Green's-function zeros manifest as an in-gap spectral weight in the unitary scattering regime. 
In this limit, we  map the impurity problem onto a doped Mott insulator and identify the resulting in-gap state as a \textit{zeron} excitation which is a \textit{localized} doublon (holon) for an attractive (repulsive) potential.
The zeron spectral weight and its associated zero vanish above a critical Zeeman field. Our results imply that Green's-function zeros have in fact already been observed in experiments, and establish impurity and magnetic-field tuning as practical tools for controlling their topology.  
\end{abstract}

\maketitle

\begin{figure*}
    \centering
    \includegraphics[width=1\linewidth]{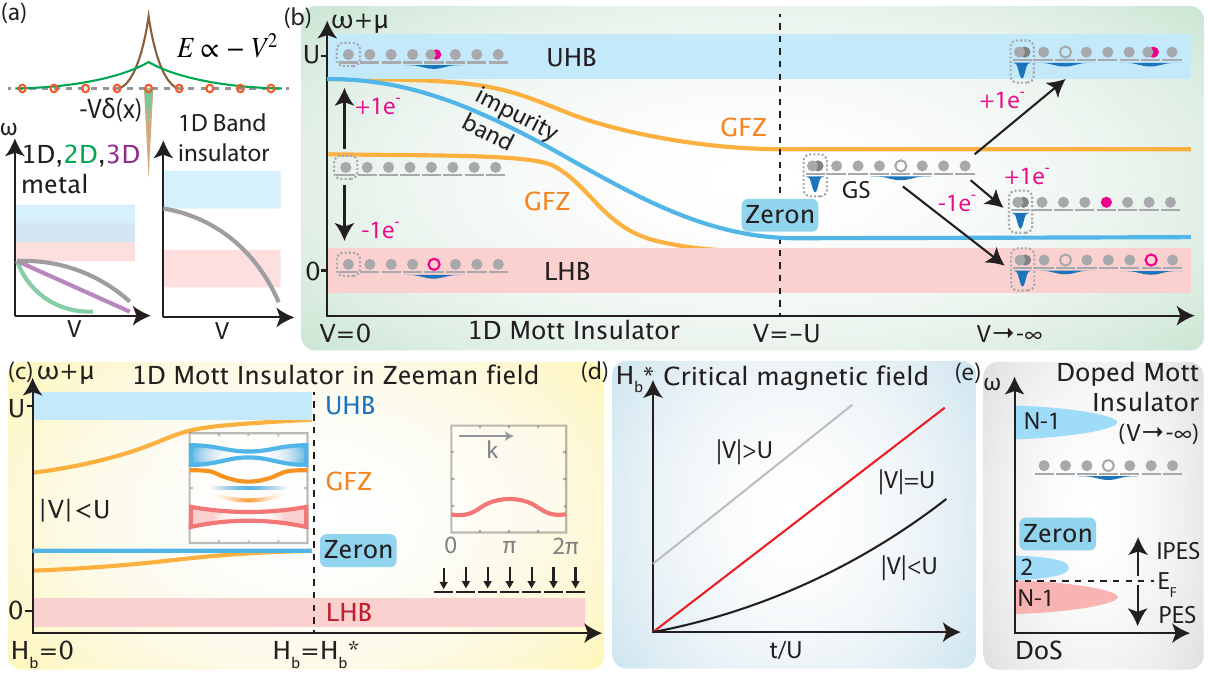}
    \caption{Schematic figure illustrating the key results: (a) variation of the impurity-induced bound state energy for a  Dirac delta impurity potential, in metals and in band insulators showing its divergence in the unitary ($V \rightarrow \infty$) limit. (b) In contrast for a Mott insulator with upper and lower Hubbard bands (UHB, LHB), and Green's function zeros (GFZs), the impurity band asymptotes for $|V|>U$ due to the GFZs forming a \textit{zeron} excitation. The disks and circles denote the content of initial and final eigenvectors leading to various observed spectral weights --  the filled circles are electron and empty are holes. The arrows indicate single particle addition $cc^\dagger$ and removal $c^\dagger c$ and the pink color indicates the added electron/hole; (c) If a Zeeman field is applied, the zerons and GFZs disappear for a critical field; the arrows indicate spin-down electrons and $k$ is the wave-vector, inset figure showing the spectral function across the Brillouin zone. (d) For $|V|<U$, the critical field can be made arbitrarily small in the unitary limit ($t/U\rightarrow 0$), making it experimentally accessible; (e) Spectral function of a single-hole–doped Mott insulator with $N$ sites, showing two in-gap zeron states above the LHB; in the unitary limit the impurity problem maps onto this doped system, identifying the zeron excitations as a manifestation of GFZs.}
    \label{fig:highlight}
\end{figure*}

\textit{Introduction}: The discovery and characterization of well-known topological phases and materials~\cite{Qi-Zhang2011, Hasan-Kane2010, Moore2010, Vishwanath2018-RMP, Bradlyn-Cano2021}
rely on the existence of well-defined quasiparticles (QPs)~\cite{Landau2013}. However, strongly correlated systems, most notably Mott insulators~\cite{Wen2006, phillips2010}, often lack such a description: their single-particle Green's function develops \textit{zeros} rather than poles~\cite{AGD2012}, signaling QP loss.
Recent theoretical work has therefore turned to the problem of defining topology in  QP-absent regimes~\cite{Si-Hu-2021,Setty-Si-2023,Fu2024} by using Green's function zeros (GFZs) as the quantities of interest~\cite{Setty-Si-2023,SettyZeros2024, SettyZeros2024-2, SettyZeros2024-3, Sangiovanni2023,  Sangiovanni2024, Si-ChenZeros2024, Sangiovanni2025, Martin2024, Fabrizio2023, Banerjee2025,Bena2025, Gurarie2011, Gurarie2011-2, Vishwanath2018, Xu2015, Volovik2003}.

Evidence from theoretical and numerical studies suggests that GFZs may carry a topological structure~\cite{Setty-Si-2023,SettyZeros2024, SettyZeros2024-2, SettyZeros2024-3, Sangiovanni2023, Sangiovanni2024, Si-ChenZeros2024, Martin2024, Fabrizio2023, Bena2025, Gurarie2011, Gurarie2011-2, Vishwanath2018, Xu2015, Volovik2003} and obey a form of bulk–boundary correspondence~\cite{SettyZeros2024-3, Sangiovanni2023, Sangiovanni2024, Martin2024,  Gurarie2011, Gurarie2011-2, Volovik2003} similar to poles, but their physical meaning and experimental status remain unsettled. Crucially, unlike poles, GFZs leave no direct signature in standard spectroscopies, raising a fundamental question of whether and how they can be detected in real materials. 
The question of whether they are, in principle, measurable, was recently addressed by some of us, who showed that GFZs contribute to observable quantities in a gauge-invariant manner consistent with physical expectations~\cite{SettyZeros2024}. Here we address the latter by showing that impurity spectroscopy provides a direct and experimentally accessible probe of GFZs and their topology through the emergence of in-gap states, which we term \textit{zeron} excitations. In Mott insulators, these zerons correspond to highly localized doublons or holons. We further argue that zeron physics has in fact already been evidenced in prior experiments~\cite{Kao1991,Chen2010, Wang2013, Wang2021-2, cai2016, Yin2022, Wen2024, Takagi2004, Zhou2021-2, Shen2001, Tajima1991,Yamada2003,Tokura1995, Moon2016, Kourkoutis2021, Hess2021, McElroy2014, Yin2022-2, Ming2017,Ming2023} in the context of doped Mott insulators. This identification is significant because it establishes a concrete link between theoretical descriptions of the topology of GFZs and their realization in measurable solid-state systems~\cite{Muechler2025,Sangiovanni2025}.

\textit{Impurity Spectroscopy:} Impurities serve as highly sensitive spectroscopic probes of electronic phases: properly tuned impurity potentials generate electron or hole bound states whose spectra encode the excitations of the surrounding host. It has been established that such impurity-induced features provide robust signatures of metals, insulators, superconductors, and correlated phases~\cite{Balatsky2006RMP}. 
We argue here that impurity spectroscopy also probes regimes of QP loss, thereby providing direct information about Green’s function zeros. A useful starting point is the standard single-site $T$-matrix framework~\cite{Balatsky2006RMP}, which captures how a local impurity hybridizes with the excitations of an interacting host~\cite{Scalapino1996,Rozenberg2008,Hirschfeld2009,Devereaux2012,Zhong2023,Si2024}. 
For an interacting system described by a Hamiltonian 
\(
H = H_I + H_{\mathrm{imp}},
\)
where \(H_I\) denotes the translationally invariant Coulomb interacting Hamiltonian and 
\(
H_{\mathrm{imp}} 
\)
represents a point-like (isotropic) impurity of strength \(V\), the single-site \(T\)-matrix formalism [Supplemental Material (SM)~\cite{SM2026} Sec. I] provides a controlled description of impurity scattering. Under simplifying assumptions of a local, momentum-independent impurity potential vertex, equivalently, \(s\)-wave scattering, the multiple-scattering processes generated by \(H_{\mathrm{imp}}\) can be resummed to give
\begin{equation}
    T(\omega) = \frac{V}{1 - V \sum_{k} G^{u}(k,\omega)},
\end{equation}
where \(G^{u}(k,\omega)\) is the  interacting Green's function of the clean system (i.e., with \(H_{\mathrm{imp}} = 0\)) 
derived typically under simplifying assumptions~\cite{Scalapino1996,Rozenberg2008,Hirschfeld2009,Devereaux2012,Zhong2023,Si2024}.
Obtaining $T(\omega)$ above requires resumming the Dyson series for the impurity Hamiltonian written in the fully interacting basis for a point impurity, and has been studied previously~\cite{Scalapino1996,Rozenberg2008,Hirschfeld2009,Devereaux2012,Zhong2023, Si2024}. \
In conventional settings like ordinary metals or band insulators where \(G^{u}(k,\omega)\) reduces to the non-interacting Green's function, the impurity–bound-state energy 
diverges in the unitary limit $V \to \infty$, with the  scaling determined by dimensionality [see schematic Fig.~\ref{fig:highlight}(a)]. 

In sharp contrast, for a Mott insulator with GFZs, the $T$-matrix in the unitary limit ($V \rightarrow \infty$) yields at least one non-divergent in-gap impurity-induced bound state 
because $\sum_{\mathbf{k}} G^{u}(k,\omega) \to 0$ in the atomic limit [Sec.~I.E of SM]. This behavior, characteristic of correlated systems lacking quasiparticles, is consistent with earlier theoretical approaches~[\citealp{Si2024},\citealp{Bena2025}] and is depicted in Fig.~\ref{fig:highlight}(b). 
We elucidate this physics using exact diagonalization of the one-dimensional Hubbard model with an onsite coulomb interaction $U$, and show that a Mott insulator hosting GFZs exhibits \emph{saturation}, rather than divergence, of the impurity bound-state energy in the unitary scattering limit. This saturated in-gap spectral weight corresponds to the emergence of a zeron. The application of a Zeeman field $H_b$ provides an additional diagnostic: above a critical field $H_b^*$, both the zeron spectral weight and the associated GFZ are quenched, as shown in Fig.~\ref{fig:highlight}(c). For $|V|\lesssim U$, $H_b^*$ increases quadratically with the ratio $t/U$, but can be made experimentally accessible in the atomic limit $t/U\ll 1$ [Fig.~\ref{fig:highlight}(d)]. Figure~\ref{fig:highlight}(e) illustrates the appearance of in-gap spectral weight in the atomic limit of a Mott insulator doped with a single hole. We show that, in the unitary limit, attractive (repulsive) impurities map onto hole- (electron-) doped Mott insulators, producing an identical spectral weight and demonstrating generically that unitary scattering produces  zeron excitations as a direct manifestation of GFZs. Our results imply that GFZs have in fact already been observed in experiments.

\textit{Model and Procedure}: The Hubbard model Hamiltonian, with a single-site non-magnetic impurity $V$ at site $l$ and Zeeman field $H_b$ is written as 

\begin{multline}
    H=\sum_{i, j} \sum_\sigma t_{i, j} c_{i, \sigma}^{\dagger} c_{j, \sigma}+U \sum_i n_{i, \uparrow} n_{i, \downarrow} \\
    + V\left(n_{l, \uparrow}+n_{l, \downarrow}\right) + H_b\left(n_{\uparrow}-n_{\downarrow}\right)
    \label{eqn:hubbard}
\end{multline}
Here, $t_{i,j} = t$ is the nearest neighbor hopping, and $c_{i, \sigma}, n_{i, \sigma}$ are the annihilation and number operator at site $i$ and spin $\sigma$.  
Exact Diagonalization (ED) is performed to evaluate the single particle Green's function [See Secs. II and III of SM]. In the presence of an impurity, the Green's function mixes momenta, even for a local isotropic potential; however, this mixing is not essential for the impurity properties of interest since the $T$-matrix itself is momentum independent for a local isotropic impurity. Further, our quantity of interest is the spectral function, as measured for example in photoemission, which requires only a single momentum label. The case of momentum mixing is discussed later and in SM. Therefore, we focus on the single-momentum Green's function, $\mathcal{G}^u_\sigma(k,\omega)$ 
given as

\begin{multline}
    \mathcal{G}^u_\sigma(k, \omega)=\left\langle\psi_0^{(N)}\right| c_{k, \sigma}^{\dagger} \frac{1}{\omega-E_0^{(N)}+ \hat{H}} c_{k, \sigma}\left|\psi_0^{(N)}\right\rangle \\
    +\left\langle\psi_0^{(N)}\right| c_{k, \sigma} \frac{1}{\omega+E_0^{(N)}-\hat{H}} c_{k, \sigma}^{\dagger}\left|\psi_0^{(N)}\right\rangle.
    \label{eqn:greens}
\end{multline}
Here $E_0^{(N)},|\psi_0^{(N)}\rangle $ are the $N$ particle ground state energy and eigenvector. The first term in Eq. \ref{eqn:greens} is the single-particle removal term, which is measured via photoemission (ARPES), and the second term is the single-particle addition, measured using inverse photoemission Spectroscopy (IPES)\cite{damascelli2003}. The the experimentally measured spectral functions are $A_\sigma(k,\omega)=-\text{Im}[\mathcal{G}^u_\sigma(k,\omega)]$, and the GFZs are identified from numerical calculations using the self-energy $\Sigma_\sigma(k,\omega)=1/G^0(k,\omega) - 1/\mathcal{G}^u_\sigma(k,\omega)$. The poles of $-\text{Im}[\Sigma](k,\omega)$ are where the GFZs are located. A broadening term $+i\eta$ with $\eta=0.05$ is added to the denominator of Eq. \ref{eqn:greens}.
Out of the 52,920 eigenstates of half-filling$\pm$1 electrons on a 10-site lattice, only 1,000 states are used for the Green's function calcuations. This number is large enough to study the range of energies around the Mott gap.
By varying $H_b$ and $V$, we calculate $\mathcal{G}_\sigma(k,\omega)$ and keep track of the poles and the zeros (GFZs) of the Green's function across the Brillouin Zone (BZ), i.e., $k=0,1\dots L\cdot2\pi/L$.

\begin{figure}[t]
\includegraphics[width=1\linewidth]{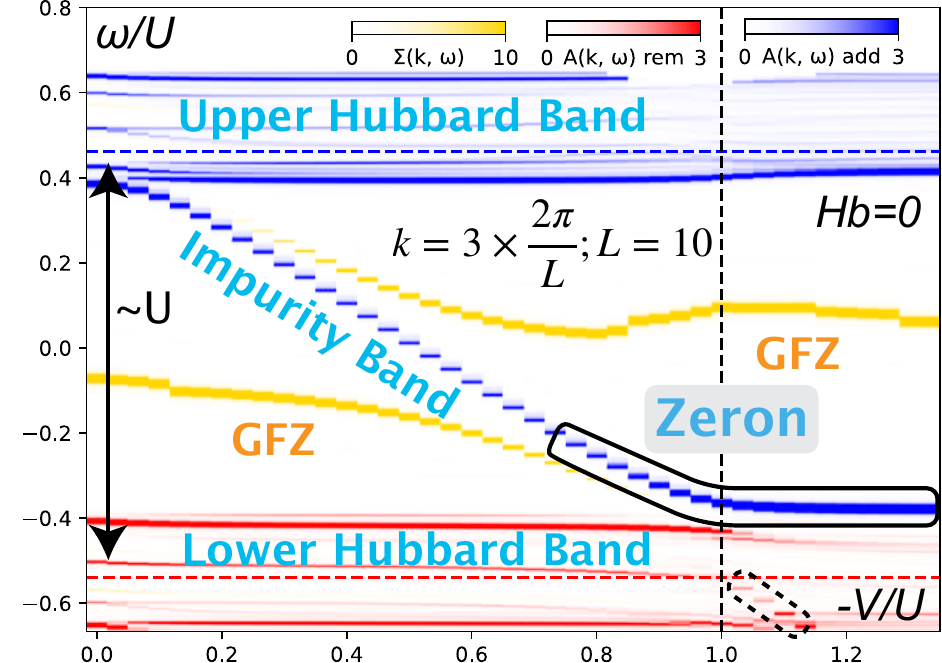}
\caption{\label{fig:spectrums_vs_V}Plot of $A_\sigma(k,\omega)$ and $\Sigma_\sigma(k,\omega)$ vs $V$ for $\sigma=\downarrow\text{ or }\uparrow, k=3/10\times 2\pi$ (no Zeeman field), calculated for a 10-site lattice, highlighting the UHB, LHB, impurity band and GFZ, and also the saturation of the impurity band above the LHB. The bands are colored based on whether they originate from the particle addition or the removal term.}
\end{figure}

\begin{figure}[b]
\raggedright
\includegraphics[width=1\linewidth]{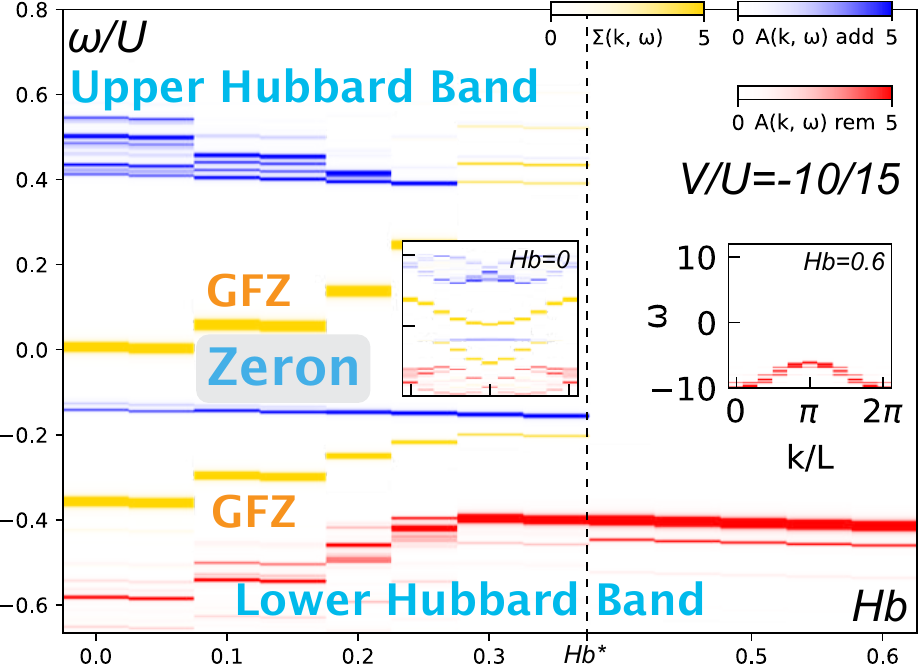}
\caption{\label{fig:Gkw_vs_Hb}Plot of $A_\sigma(k,\omega)$ and $\Sigma_\sigma(k,\omega)$ vs $H_b$ for $\sigma=\downarrow, k=5/10\times 2\pi$, and $V/U=-10/15$. The UHB, LHB, and the impurity band are highlighted, showing the disappearance of impurity band for $H_b>H_b^*$. The inset plots are across the BZ for $H_b=0$ and $H_b=0.6>H_b^*$.}
\end{figure}

\begin{figure}[b]
\includegraphics[width=1\linewidth]{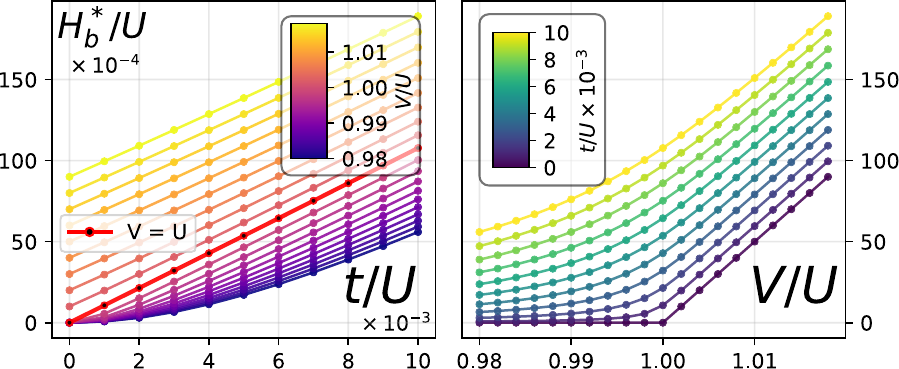}
\caption{\label{fig:Hb_crit} $H_b^*/U$ vs (a)$t/U$ and (b)$V/U$ in the unitarity limit ($U/t\rightarrow\infty$). $H_b^*$ is parabolic on $t$ for $V<U$ and linear for $V>U$. In the unitarity limit,$H_b^*$ is nearly constant close to zero for $V<U$ and linear for $V>U$.}
\end{figure}

\textit{Results}: In Fig. \ref{fig:spectrums_vs_V}, the spectral functions $A_\sigma(k,\omega)$ and $\Sigma_\sigma(k,\omega)$ are calculated for a 10-site lattice and plotted for $k=3\times\frac{2\pi}{L}$. As described in Eq. \ref{eqn:greens}, the Green's function is written as the sum of a particle addition and a particle removal term. The spin-up or down particle addition term, labeled as ``$A(k,\omega)\text{ add}$" is plotted in blue, and particle removal term, labeled as ``$A(k,\omega)\text{ rem}$", is plotted in red. At zero magnetic field ($H_b=0$), as we introduce a small, non-zero $V$, an additional band, which we identify as the \textit{impurity band} [Fig. \ref{fig:spectrums_vs_V}], splits off from the UHB for $V<0$ and LHB for $V>0$. As $|V|$ is increased from 0 to $U$, the impurity band travels in the Mott gap, asymptoting near the LHB for $V\approx -U$ or the UHB for $V\approx U$, \emph{indicating an additional in-gap bound state}, in the unitary limit. We will see below that this additional spectral weight corresponds to a highly localized doublon (holon) for an attractive (repulsive) potential. As discussed in the main text, we identify this excitation as the \textit{zeron} that results from the presence of GFZs in the parent Mott insulator.

To understand the nature of the zeron excitation, we inspect the eigenvectors in large impurity limit.  We observe that the ground state $\psi^0_{N}$ has a localized doublon in the impurity site and a holon delocalized on the non-impurity sites [Fig.~\ref{fig:highlight}(b) and SM Sec. IV]. In order to understand why the degeneracy in the UHB is broken for $|V|>U$, one can look at the single particle addition ground state $\psi^0_{(N+1)}$ and excited state $\psi^{\text{excited}}_{(N+1)}$, in particular the expection value of the site-wise number operator. The ground state $\psi^0_{(N+1)}$ has a localized doublon without any delocalized holon, and the excited state $\psi^{\text{excited}}_{(N+1)}$ has a localized doublon, a delocalized holon and a delocalized doublon. These particle addition states lead to the zeron and the LHB respectively. For no impurity, $\psi^0_{(N+1)}$ and $\psi^{\text{excited}}_{(N+1)}$ are degenerate because there is no localized holon or doublon [SM Sec.~IV]. The particle removal states $\psi_{(N-1)}$ around the Mott gap are still degenerate for any value of impurity.
For $V\rightarrow \infty$, the impurity site has a localized holon [SM Sec.~IV] and the degeneracy is broken in the LHB for the particle removal states. A key conclusion of our eigenstate analysis is that the impurity site effectively decouples from the system in the unitary limit, making it one-electron doped ($V\rightarrow\infty$) or one-hole doped ($V\rightarrow -\infty$)[SM Sec.~IV]. Thus the analytic behavior of the Hubbard model with large on-site potential can be mapped to doping a Mott insulator.

 The latter has been extensively investigated previously and an exact analytical treatment in the atomic limit is given in Refs.~\cite{Wen2006, phillips2010, meinders1993}. In the one-hole doped case, the missing electron shifts spectral weight between the lower and upper Hubbard bands while preserving the total of two states per site. The doped hole removes one state from the UHB and adds one to the LHB so that their combined weight remains fixed, and the empty impurity site contributes two low-energy addition states (spin up or spin down) inside the Mott gap, as shown in Fig.~\ref{fig:highlight}(e). Consequently, the impurity-induced in-gap band saturates at exactly the same total in-gap spectral weight as expected for a doped Mott insulator, providing a simple counting picture for the impurity-doped system \cite{meinders1993,phillips2010}. \emph{Hence, we conclude that the in-gap spectral weight for doped Mott insulator can be interpreted as a zeron for Hubbard model with a potential impurity}.
In contrast, for a band insulator where 
no GFZ occurs, 
a mid-band is seen for non-zero $V$, but it does not asymptote within the gap.  By removing Coloumb interaction, $U=0$, one can also study a single band metal, in which case also no GFZs or zerons are observed [See SM Sec. V].

We now study the effect of a Zeeman  field by varying the $H_b$ term in the Hamiltonian (Eq. \ref{eqn:hubbard}). For no impurity ($V=0$), as $H_b$ is increased from zero, the system gets polarized until at a critical $H_b$, herein referred to as $H_b^*$, the system gets fully polarized $\left<S_z\right> = -L/2$. Numerically, $H_b^*$ is obtained where the $\left<S_z\right> = -L/2$ ground state energy crosses the $\left<S_z\right> = -L/2+1$ ground state energy [See SM Sec.~VI.B].
For $H_b>H_b^*>0$, the system is fully polarized with spin-down electrons, and with spin-up electrons for $H_b<H_b^*<0$ [See SM Sec.~VI.A,C]. In this fully polarized limit the model is exactly solvable and has been extensively studied, including within dynamical mean-field theory~\cite{denteneer2003}. Because the half-filled system contains electrons of a single spin species, there are no available states to add an electron of the same spin, and the UHB and GFZ are therefore expected to disappear. As a result, the zeron excitation is also expected to be quenched in the fully polarized phase.
This behavior is also expected for non-zero $V$ and $|V|<U$.  In Fig. \ref{fig:Gkw_vs_Hb}, the spin down spectral functions are plotted vs $H_b$ in the center of BZ for $V/U=10/15$ [see SM Sec.~VI B for a discussion on spin asymmetry]. Notice the disappearance of the zeron, UHB, and GFZ for $H_b>H_b^*$. The inset panels show the spectral functions across BZ for $H_b=0$ and $H_b=0.6>H_b^*$. Detailed studies on the stability of GFZs to interactions appear in Refs.~\cite{Setty-Luttinger2020, Setty-Luttinger2021, Setty2021-2, PWP2022}. 

\begin{figure}[b]
    \centering
    \includegraphics[width=1\linewidth]{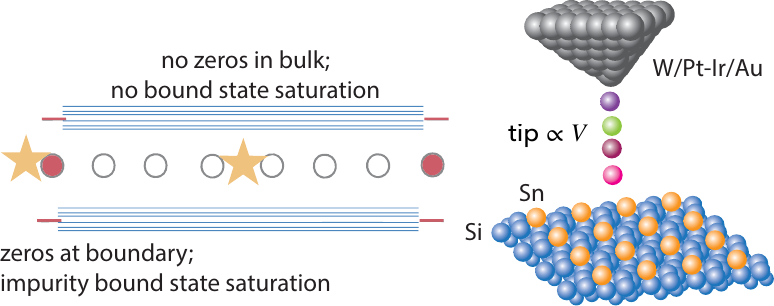}
    \caption{Probing boundary GFZs: (Left) Projection of bulk-boundary states with impurity. As the bulk is an ordinary insulator without GFZs while the boundary has GFZs associated with interaction-gapped edge states, zerons occur only when the impurity is localized at the boundary. (Right) Proposed STM/STS experiment in Mott-insulating materials with engineered dopants added on Sn/Si(111)~\cite{Weitering2017, Weitering2023}, where the tip is used to tune the impurity potential.  
    }
    \label{fig:exp}
\end{figure}

We now show that $H_b^*$ may be experimentally accessible in certain limiting cases [see also SM Sec.~VI.C].  
Fig.~\ref{fig:Hb_crit}(a) shows $H_b^*/U$ versus $t/U$ for different $V$, illustrating that for $V\lesssim U$ the critical field is quadratic in $t/U$ and vanishes in the atomic limit $t/U\to0$. The dependence of $H_b^*$ with $t/U$ crosses over from quadratic to linear as $V$ increases, and for $V\gtrsim U$, $H_b^*$ becomes linear. In this regime, full spin polarization requires the Zeeman field to energetically decouple the doublon localized at the impurity site, whose potential energy is set by $V$, leading to the observed linear scaling of $H_b^*$ [see SM Sec.~VI.C].  Fig.~\ref{fig:Hb_crit}(b) plots $H_b^*/U$ as a function of $V/U$ for several values of $t/U$; notably, in the atomic limit $H_b^*$ can be made arbitrarily small for $V\lesssim U$. 
Finally, we note that for inelastic scattering, characterized by $A_\sigma(k,k'\neq k,\omega)$ [see SM Sec.~VIII], the upper and lower Hubbard bands are strongly suppressed, while the zeron contribution remains pronounced for $k\approx k'$.
To realize a minimal setup to diagnose the topology of GFZs, Fig.~\ref{fig:exp} illustrates the interacting topological systems with boundary GFZs. In the noninteracting model, the bulk is gapped while topological boundary modes reside at the edges without any GFZs. Upon introducing Coulomb interactions (ideally smaller than the bulk gap), the boundary modes gap out into GFZs pinned to the boundary, whereas the bulk remains free of GFZs. Consequently, impurities in the boundary (bulk) lead to occurrence (absence) of zerons.
This behavior establishes how impurity response can offer a diagnostic of the topology of GFZs.

\textit{Materials and experimental considerations.}
As discussed above, in the unitary limit of impurity scattering the problem of a potential impurity in a Mott insulator becomes equivalent to that of a doped Mott insulator. In this limit, several experiments have already reported the emergence of in-gap spectral weight---the very feature predicted by our theory. In the cuprate context, oxygen K-edge X-ray absorption spectroscopy on La$_{2-x}$Sr$_x$CuO$_{4+\delta}$ and YBa$_2$Cu$_3$O$_{6+x}$ demonstrates doping-dependent appearance of in-gap states~\cite{Kao1991,Chen2010}. STM/STS on Ca$_2$CuO$_2$Cl$_2$ reveals particle--hole symmetric Hubbard bands and strongly localized in-gap states around intrinsic defects~\cite{Wang2013}. Controlled electron doping of Ca$_2$CuO$_2$Cl$_2$ by sub-monolayer Rb deposition further demonstrates progressive Mott-gap filling and emergent in-gap spectral  weight~\cite{Wen2024}, with related STM observations in (Ca,Na)$_2$CuO$_2$Cl$_2$~\cite{Takagi2004}. Momentum-resolved ARPES using sequential in-situ electron doping reveals spectral-weight transfer and the appearance of low-energy states in the Mott gap~\cite{Zhou2021-2, Shen2001}. Optical and infrared spectroscopy show doping-driven redistribution of spectral weight in La$_{2-x}$Sr$_x$CuO$_4$~\cite{Tajima1991,Yamada2003,Tokura1995} and in electron-doped (Sr$_{1-x}$La$_x$)$_3$Ir$_2$O$_7$~\cite{Moon2016}.  STEM--EELS on Nd$_{1-x}$Sr$_x$NiO$_2$ further reveals a reshaping of the Mott--Hubbard landscape with carrier addition~\cite{Kourkoutis2021}. Related impurity-induced in-gap states have also been observed in oxygen-deficient Sr$_2$IrO$_4$~\cite{Hess2021,McElroy2014}, in 1T--TaS$_2$ and related dichalcogenides~\cite{Yin2022-2}, and in the hole-doped monolayer Sn/Si(111) system showing Mottness~\cite{Weitering2017, Weitering2023}. This impurity or doping induced in-gap spectral weight provides support for zeron excitations.

What is missing, however, are experiments where the impurity strength is systematically tuned from weak to unitary scattering, and thereby track the evolution and saturation of the impurity-bound-state (zeron) energy as the unitary limit is approached. We therefore propose STM/STS experiments in which dilute, nominally nonmagnetic impurities---substitutional dopants, vacancies, or adatoms---are intentionally introduced into correlated insulators such as Sr$_2$IrO$_4$, Ca$_2$CuO$_2$Cl$_2$, 1T--TaS$_2$, Sn/Si(111) or suitable pyrite and Kondo insulators.  By imaging the local density of states around individual impurities while varying impurity species [Fig.~\ref{fig:exp}(right)], local environments, or tip-induced potentials, one can follow the bound-state dispersion as the effective scattering strength is driven toward the Mott scale. The atomic manipulation capabilities of STM~\cite{Ko2019} further enable systematic construction of impurity configurations, allowing in-situ verification of the predicted saturation of the impurity band. Our theory predicts that, in the presence of GFZs, this impurity band approaches a finite zeron excitation energy rather than diverging, and that both the zeron band and the GFZ are suppressed above a critical magnetic field $H_b^*$, which lies within the field range of modern STM experiments.

\textit{Discussion and conclusions:} Several points are in order. First, we stress that the GFZs studied here are interaction-driven, arising from Coulomb correlations that generate zeros in the determinant of the Green's function in a multi-orbital system. This is fundamentally distinct from the noninteracting, topology-induced zeros appearing in individual matrix elements in the orbital basis~\cite{Yamaji2022,Balents2015,Stern2024,Hosur2025}, which are not the focus of this work. Second, although the critical field $H_b^*$ can be made arbitrarily small for $V/U<1$ near the atomic limit, it rapidly becomes impractically large away from this regime as it scales with $U$. This motivates targeting $f$-electron systems and Kondo insulators, where $H_b^*$ is more accessible and the unitary limit is easier to achieve. Finally, our results are consistent with Refs.~[\citealp{Si2024},\citealp{Bena2025}] which report properties of in-gap impurity states in a Mott insulator within  slave-rotor and Hubbard operator approaches.

To conclude, we have shown that GFZs in Mott insulators leave experimentally accessible signatures through impurity spectroscopy. Using exact diagonalization of the Hubbard model with a potential impurity and Zeeman field, supported by analytic arguments, we demonstrated that a GFZ inside the Mott gap produces a saturated impurity band in the unitary scattering limit that we term as a \textit{zeron} excitation. In the Mott context, zerons are highly localized doublons (holons) nucleated by a strong attractive (repulsive) impurity potential. This is in sharp contrast to the divergent behavior expected in conventional band insulators and metals. By mapping the unitary impurity problem onto a doped Mott insulator, we identified zerons in well-known Mott insulators, implying that GFZs have in fact already been realized experimentally. We further showed that both the zeron and associated GFZs vanish above a critical Zeeman field, providing a tunable and falsifiable diagnostic of quasiparticle breakdown. Finally, we proposed a platform in which the impurity strength can be tuned toward the unitary limit, enabling direct observation of zerons and its magnetic-field control. Together, these results elevate impurity scattering to a quantitative experimental tool for detecting and manipulating topology beyond QPs.

\textit{Acknowledgements:} We thank M. Tringides, R. Flint, W. Ding, P. Canfield, A. Kaminski and R. Prozorov for discussions. S.M and C.S are supported by Iowa State University start-up funds and Laboratory Directed Research and Development program at Ames Lab 
with the U.S. Department of Energy.  Work at Rice has been supported by NSF Grant No. DMR-2220603 (F.X.), and by the Robert A. Welch Foundation Grant No. C-1411 and VBFF ONR-VB N00014-23-1-2870 (Q.S.). M.K. was supported by the DOE Early Career Research Program. Part of the research was performed at the Ames National Laboratory which is operated for the USDOE by Iowa State University under Contract No. DE-C02-07CH11358.  \\ \newline
\noindent
${\dagger}$ Email for correspondence: csetty@iastate.edu
\bibliography{References}

@article{Si-Hu-2021,
  title={Topological semimetals without quasiparticles},
  author={Hu, Haoyu and Chen, Lei and Setty, Chandan and Garcia-Diez, Mikel and Grefe, Sarah E and Prokofiev, Andrey and Kirchner, Stefan and Vergniory, Maia G and Paschen, Silke and Cano, Jennifer and others},
  journal={arXiv preprint arXiv:2110.06182},
  year={2021}
}

@article{Yin2022,
  title={Correlation-corrected band topology and topological surface states in iron-based superconductors},
  author={Ma, Xiaobo and Wang, Guangwei and Liu, Rui and Yu, Tianye and Peng, Yiran and Zheng, Pengyu and Yin, Zhiping},
  journal={Physical Review B},
  volume={106},
  number={11},
  pages={115114},
  year={2022},
  publisher={APS}
}

@article{Setty-Si-2023,
  title={Topological diagnosis of strongly correlated electron systems},
  author={Setty, Chandan and Xie, Fang and Sur, Shouvik and Chen, Lei and Paschen, Silke and Vergniory, Maia G and Cano, Jennifer and Si, Qimiao},
  journal={arXiv preprint arXiv:2311.12031},
  year={2023}
}

@article{SettyZeros2024,
  title={Electronic properties, correlated topology, and Green's function zeros},
  author={Setty, Chandan and Xie, Fang and Sur, Shouvik and Chen, Lei and Vergniory, Maia G and Si, Qimiao},
  journal={Physical Review Research},
  volume={6},
  number={3},
  pages={033235},
  year={2024},
  publisher={APS}
}

@article{SettyZeros2024-2,
  title={Symmetry constraints and spectral crossing in a Mott insulator with Green's function zeros},
  author={Setty, Chandan and Sur, Shouvik and Chen, Lei and Xie, Fang and Hu, Haoyu and Paschen, Silke and Cano, Jennifer and Si, Qimiao},
  journal={Physical Review Research},
  volume={6},
  number={3},
  pages={L032018},
  year={2024},
  publisher={APS}
}

@article{SettyZeros2024-3,
  title={Topological Green’s function zeros in an exactly solved model and beyond},
  author={Bollmann, Steffen and Setty, Chandan and Seifert, Urban FP and K{\"o}nig, Elio J},
  journal={Physical Review Letters},
  volume={133},
  number={13},
  pages={136504},
  year={2024},
  publisher={APS}
}

@article{phillips2010,
  title={Colloquium: Identifying the propagating charge modes in doped Mott insulators},
  author={Phillips, Philip},
  journal={Reviews of Modern Physics},
  volume={82},
  number={2},
  pages={1719--1742},
  year={2010},
  publisher={APS}
}

@article{Fabrizio2023,
  title={Unified role of Green's function poles and zeros in correlated topological insulators},
  author={Blason, Andrea and Fabrizio, Michele},
  journal={Physical Review B},
  volume={108},
  number={12},
  pages={125115},
  year={2023},
  publisher={APS}
}

@book{Volovik2003,
  title={The universe in a helium droplet},
  author={Volovik, Grigory E},
  volume={117},
  year={2003},
  publisher={OUP Oxford}
}

@article{Setty-Luttinger2021,
  title={Superconductivity from Luttinger surfaces: Emergent Sachdev-Ye-Kitaev physics with infinite-body interactions},
  author={Setty, Chandan},
  journal={Physical Review B},
  volume={103},
  number={1},
  pages={014501},
  year={2021},
  publisher={APS}
}

@article{Setty-Luttinger2020,
  title={Pairing instability on a Luttinger surface: A non-Fermi liquid to superconductor transition and its Sachdev-Ye-Kitaev dual},
  author={Setty, Chandan},
  journal={Physical Review B},
  volume={101},
  number={18},
  pages={184506},
  year={2020},
  publisher={APS}
}

@article{Balents2015,
  title={Impurity-bound states and Green's function zeros as local signatures of topology},
  author={Slager, Robert-Jan and Rademaker, Louk and Zaanen, Jan and Balents, Leon},
  journal={Physical Review B},
  volume={92},
  number={8},
  pages={085126},
  year={2015},
  publisher={APS}
}

@article{Balatsky2006RMP,
  title={Impurity-induced states in conventional and unconventional superconductors},
  author={Balatsky, Alexander V and Vekhter, Ilya and Zhu, Jian-Xin},
  journal={Reviews of Modern Physics},
  volume={78},
  number={2},
  pages={373--433},
  year={2006},
  publisher={APS}
}

@article{Sangiovanni2023,
  title={Mott insulators with boundary zeros},
  author={Wagner, Niklas and Crippa, Lorenzo and Amaricci, Adriano and Hansmann, Philipp and Klett, Marcel and K{\"o}nig, EJ and Sch{\"a}fer, Thomas and Sante, D Di and Cano, Jennifer and Millis, AJ and others},
  journal={Nature Communications},
  volume={14},
  number={1},
  pages={7531},
  year={2023},
  publisher={Nature Publishing Group UK London}
}

@article{Sangiovanni2024,
  title={Edge zeros and boundary spinons in topological Mott insulators},
  author={Wagner, Niklas and Guerci, Daniele and Millis, Andrew J and Sangiovanni, Giorgio},
  journal={Physical Review Letters},
  volume={133},
  number={12},
  pages={126504},
  year={2024},
  publisher={APS}
}

@article{Sangiovanni2025,
  title={Probing Green's Function Zeros by Co-tunneling through Mott Insulators},
  author={Lehmann, Carl and Crippa, Lorenzo and Sangiovanni, Giorgio and Budich, Jan Carl},
  journal={arXiv preprint arXiv:2502.19479},
  year={2025}
}

@article{Si-ChenZeros2024,
  title={Dirac zeros in an orbital selective Mott phase: Green's function Berry curvature and flux quantization},
  author={Chen, Lei and Hu, Haoyu and Vergniory, Maia G and Cano, Jennifer and Si, Qimiao},
  journal={arXiv preprint arXiv:2401.12156},
  year={2024}
}

@article{Martin2024,
  title={Global anomalies of Green's function zeros},
  author={Su, Lei and Martin, Ivar},
  journal={arXiv preprint arXiv:2405.08093},
  year={2024}
}

@article{Vishwanath2018,
  title={Symmetric fermion mass generation as deconfined quantum criticality},
  author={You, Yi-Zhuang and He, Yin-Chen and Xu, Cenke and Vishwanath, Ashvin},
  journal={Physical Review X},
  volume={8},
  number={1},
  pages={011026},
  year={2018},
  publisher={APS}
}

@article{Xu2015,
  title={Exotic quantum phase transitions of strongly interacting topological insulators},
  author={Slagle, Kevin and You, Yi-Zhuang and Xu, Cenke},
  journal={Physical Review B},
  volume={91},
  number={11},
  pages={115121},
  year={2015},
  publisher={APS}
}

@article{Muechler2025,
  title={Strong Correlations, Green's Function Zeros and Topological Transitions in Orbital-Symmetry-Controlled Chemical Reactions},
  author={Xie, Ziren and Mirzanejad, Amir and Muechler, Lukas},
  journal={arXiv preprint arXiv:2506.18984},
  year={2025}
}

@book{Landau2013,
  title={Statistical Physics: Volume 5},
  author={Landau, Lev Davidovich and Lifshitz, Evgenii Mikhailovich},
  volume={5},
  year={2013},
  publisher={Elsevier}
}

@article{Wen2006,
  title={Doping a Mott insulator: Physics of high-temperature superconductivity},
  author={Lee, Patrick A and Nagaosa, Naoto and Wen, Xiao-Gang},
  journal={Reviews of modern physics},
  volume={78},
  number={1},
  pages={17--85},
  year={2006},
  publisher={APS}
}

@article{Gurarie2011,
  title={Single-particle Green’s functions and interacting topological insulators},
  author={Gurarie, Victor},
  journal={Physical Review B—Condensed Matter and Materials Physics},
  volume={83},
  number={8},
  pages={085426},
  year={2011},
  publisher={APS}
}

@article{Gurarie2011-2,
  title={Bulk-boundary correspondence of topological insulators from their respective Green’s functions},
  author={Essin, Andrew M and Gurarie, Victor},
  journal={Physical Review B—Condensed Matter and Materials Physics},
  volume={84},
  number={12},
  pages={125132},
  year={2011},
  publisher={APS}
}

@book{AGD2012,
  title={Methods of quantum field theory in statistical physics},
  author={Abrikosov, Aleksei Alekseevich and Gorkov, Lev Petrovich and Dzyaloshinski, Igor Ekhielevich},
  year={2012},
  publisher={Courier Corporation}
}

@article{Rozenberg2008,
  title={Impurity scattering in a strongly correlated host},
  author={Lederer, P and Rozenberg, Marcelo Javier},
  journal={Europhysics Letters},
  volume={81},
  number={6},
  pages={67002},
  year={2008},
  publisher={IOP Publishing}
}

@article{Scalapino1996,
  title={T-matrix formulation of impurity scattering in correlated systems},
  author={Ziegler, W and Poilblanc, D and Preuss, R and Hanke, W and Scalapino, DJ},
  journal={Physical Review B},
  volume={53},
  number={13},
  pages={8704},
  year={1996},
  publisher={APS}
}

@article{Zhong2023,
  title={Friedel oscillation in non-Fermi liquid: lesson from exactly solvable Hatsugai--Kohmoto model},
  author={Zhao, Miaomiao and Yang, Wei-Wei and Luo, Hong-Gang and Zhong, Yin},
  journal={Journal of Physics: Condensed Matter},
  volume={35},
  number={49},
  pages={495603},
  year={2023},
  publisher={IOP Publishing}
}

@article{Devereaux2012,
  title={Quasiparticle interference and the interplay between superconductivity and density wave order in the cuprates},
  author={Nowadnick, EA and Moritz, B and Devereaux, TP},
  journal={Physical Review B—Condensed Matter and Materials Physics},
  volume={86},
  number={13},
  pages={134509},
  year={2012},
  publisher={APS}
}

@article{Hirschfeld2009,
  title={Defects in correlated metals and superconductors},
  author={Alloul, H and Bobroff, J and Gabay, M and Hirschfeld, PJ},
  journal={Reviews of Modern Physics},
  volume={81},
  number={1},
  pages={45--108},
  year={2009},
  publisher={APS}
}

@article{Si2024,
  title={Local density of states induced near impurities in Mott insulators},
  author={Ding, Wenxin and Si, Qimiao},
  journal={Physical Review B},
  volume={110},
  number={8},
  pages={L081104},
  year={2024},
  publisher={APS}
}

@article{Fu2024,
  title={Non-Hermitian topological theory of finite-lifetime quasiparticles: Prediction of bulk Fermi arc due to exceptional point},
  author={Kozii, Vladyslav and Fu, Liang},
  journal={Physical Review B},
  volume={109},
  number={23},
  pages={235139},
  year={2024},
  publisher={APS}
}

@article{cai2016,
  title={Visualizing the evolution from the Mott insulator to a charge-ordered insulator in lightly doped cuprates},
  author={Cai, Peng and Ruan, Wei and Peng, Yingying and Ye, Cun and Li, Xintong and Hao, Zhenqi and Zhou, Xingjiang and Lee, Dung-Hai and Wang, Yayu},
  journal={Nature Physics},
  volume={12},
  number={11},
  pages={1047--1051},
  year={2016},
  publisher={Nature Publishing Group UK London}
}

@article{meinders1993,
  title = {Spectral-weight transfer: Breakdown of low-energy-scale sum rules in correlated systems},
  author = {Meinders, M. B. J. and Eskes, H. and Sawatzky, G. A.},
  journal = {Phys. Rev. B},
  volume = {48},
  issue = {6},
  pages = {3916--3926},
  numpages = {0},
  year = {1993},
  month = {Aug},
  publisher = {American Physical Society},
  doi = {10.1103/PhysRevB.48.3916},
  url = {https://link.aps.org/doi/10.1103/PhysRevB.48.3916}
}

@article{denteneer2003,
  title = {Interacting Electrons in a Two-Dimensional Disordered Environment: Effect of a Zeeman Magnetic Field},
  author = {Denteneer, P. J. H. and Scalettar, R. T.},
  journal = {Phys. Rev. Lett.},
  volume = {90},
  issue = {24},
  pages = {246401},
  numpages = {4},
  year = {2003},
  month = {Jun},
  publisher = {American Physical Society},
  doi = {10.1103/PhysRevLett.90.246401},
  url = {https://link.aps.org/doi/10.1103/PhysRevLett.90.246401}
}

@article{damascelli2003,
  title={Angle-resolved photoemission studies of the cuprate superconductors},
  author={Damascelli, Andrea and Hussain, Zahid and Shen, Zhi-Xun},
  journal={Reviews of modern physics},
  volume={75},
  number={2},
  pages={473},
  year={2003},
  publisher={APS}
}

@article{Hess2021,
  title={Evidence for a percolative Mott insulator-metal transition in doped Sr 2 IrO 4},
  author={Sun, Zhixiang and Guevara, Jose M and Sykora, Steffen and P{\"a}rschke, Ekaterina M and Manna, Kaustuv and Maljuk, Andrey and Wurmehl, Sabine and Van Den Brink, Jeroen and B{\"u}chner, Bernd and Hess, Christian},
  journal={Physical Review Research},
  volume={3},
  number={2},
  pages={023075},
  year={2021},
  publisher={APS}
}

@article{McElroy2014,
  title={Local density of states study of a spin-orbit-coupling induced Mott insulator Sr 2 Ir O 4},
  author={Dai, Jixia and Calleja, Eduardo and Cao, Gang and McElroy, Kyle},
  journal={Physical Review B},
  volume={90},
  number={4},
  pages={041102},
  year={2014},
  publisher={APS}
}

@article{Wang2021-2,
  title={Imaging the atomic-scale electronic states induced by a pair of hole dopants in Ca2CuO2Cl2 Mott insulator},
  author={Li, Haiwei and Ye, Shusen and Zhao, Jianfa and Jin, Changqing and Wang, Yayu},
  journal={Science Bulletin},
  volume={66},
  number={14},
  pages={1395--1400},
  year={2021},
  publisher={Elsevier}
}

@article{Wang2013,
  title={Visualizing the atomic-scale electronic structure of the Ca2CuO2Cl2 Mott insulator},
  author={Ye, Cun and Cai, Peng and Yu, Runze and Zhou, Xiaodong and Ruan, Wei and Liu, Qingqing and Jin, Changqing and Wang, Yayu},
  journal={Nature communications},
  volume={4},
  number={1},
  pages={1365},
  year={2013},
  publisher={Nature Publishing Group UK London}
}

@article{Yin2022-2,
  title={Visualizing the evolution from mott insulator to anderson insulator in Ti-doped 1 T-TaS2},
  author={Zhang, Wenhao and Gao, Jingjing and Cheng, Li and Bu, Kunliang and Wu, Zongxiu and Fei, Ying and Zheng, Yuan and Wang, Li and Li, Fangsen and Luo, Xuan and others},
  journal={npj Quantum Materials},
  volume={7},
  number={1},
  pages={8},
  year={2022},
  publisher={Nature Publishing Group UK London}
}

@article{Wen2024,
  author       = {Han Li and Zhaohui Wang and Shengtai Fan and Huazhou Li and Huan Yang and Haihu Wen},
  title        = {Mott Gap Filling by Doping Electrons through Depositing One Sub-Monolayer Thin Film of Rb on Ca$_{2}$CuO$_{2}$Cl$_{2}$},
  journal      = {Chinese Physics Letters},
  volume       = {41},
  number       = {5},
  pages        = {057402},
  year         = {2024},
  doi          = {10.1088/0256-307X/41/5/057402}
}

@article{Zhou2021-2,
  title={Momentum-resolved visualization of electronic evolution in doping a Mott insulator},
  author={Hu, Cheng and Zhao, Jianfa and Gao, Qiang and Yan, Hongtao and Rong, Hongtao and Huang, Jianwei and Liu, Jing and Cai, Yongqing and Li, Cong and Chen, Hao and others},
  journal={Nature Communications},
  volume={12},
  number={1},
  pages={1356},
  year={2021},
  publisher={Nature Publishing Group UK London}
}

@article{Tajima1991,
  title={Optical spectra of La 2- x Sr x CuO 4: Effect of carrier doping on the electronic structure of the CuO 2 plane},
  author={Uchida, S and Ido, T and Takagi, H and Arima, T and Tokura, Y and Tajima, S},
  journal={Physical Review B},
  volume={43},
  number={10},
  pages={7942},
  year={1991},
  publisher={APS}
}

@article{Moon2016,
  title={Infrared Spectroscopic Evidences of Strong Electronic Correlations in (Sr1- x La x) 3Ir2O7},
  author={Ahn, Gihyeon and Song, SJ and Hogan, T and Wilson, SD and Moon, SJ},
  journal={Scientific reports},
  volume={6},
  number={1},
  pages={32632},
  year={2016},
  publisher={Nature Publishing Group UK London}
}

@article{Kourkoutis2021,
  title={Doping evolution of the Mott--Hubbard landscape in infinite-layer nickelates},
  author={Goodge, Berit H and Li, Danfeng and Lee, Kyuho and Osada, Motoki and Wang, Bai Yang and Sawatzky, George A and Hwang, Harold Y and Kourkoutis, Lena F},
  journal={Proceedings of the National Academy of Sciences},
  volume={118},
  number={2},
  pages={e2007683118},
  year={2021},
  publisher={National Academy of Sciences}
}

@article{Chen2010,
  title={Dynamical spectral weight in YBa $ \_2 $ Cu $ \_3 $ O $ \_y $ probed by x-ray absorption spectroscopy},
  author={Lin, J-Y and Lee, PR and Liu, YT and Mou, Chung-Yu and Chen, Y-J and Wu, KH and Luo, CW and Juang, JY and Uen, TM and Lee, JM and others},
  journal={arXiv preprint arXiv:1009.2560},
  year={2010}
}

@article{Kao1991,
  title={Electronic states in La 2- x Sr x CuO 4+ $\delta$ probed by soft-x-ray absorption},
  author={Chen, CT and Sette, F and Ma, Y and Hybertsen, MS and Stechel, EB and Foulkes, WMC and Schulter, M and Cheong, Sang-Wook and Cooper, AS and Rupp Jr, LW and others},
  journal={Physical review letters},
  volume={66},
  number={1},
  pages={104},
  year={1991},
  publisher={APS}
}

@article{Yamada2003,
  title={Phase Diagram of L a 2-x S rx C u O 4 Probed in the Infared: Imprints of Charge Stripe Excitations},
  author={Lucarelli, A and Lupi, Stefano and Ortolani, Michele and Calvani, Paolo and Maselli, P and Capizzi, M and Giura, P and Eisaki, H and Kikugawa, N and Fujita, T and others},
  journal={Physical review letters},
  volume={90},
  number={3},
  pages={037002},
  year={2003},
  publisher={APS}
}

@article{Tokura1995,
  title={Spectral weight transfer of the optical conductivity in doped Mott insulators},
  author={Katsufuji, T and Okimoto, Y and Tokura, Y},
  journal={Physical review letters},
  volume={75},
  number={19},
  pages={3497},
  year={1995},
  publisher={APS}
}

@article{Shen2001,
  title={From Mott insulator to overdoped superconductor: evolution of the electronic structure of cuprates studied by ARPES},
  author={Damascelli, A and Lu, DH and Shen, Z-X},
  journal={Journal of Electron Spectroscopy and Related Phenomena},
  volume={117},
  pages={165--187},
  year={2001},
  publisher={Elsevier}
}

@article{Takagi2004,
  title={Imaging Nanoscale Electronic Inhomogeneity in the Lightly Doped Mott Insulator C a 2-x N ax C u O 2 C l 2},
  author={Kohsaka, Y and Iwaya, K and Satow, S and Hanaguri, T and Azuma, M and Takano, M and Takagi, H},
  journal={Physical review letters},
  volume={93},
  number={9},
  pages={097004},
  year={2004},
  publisher={APS}
}

@article{Ming2017,
  title = {Realization of a Hole-Doped Mott Insulator on a Triangular Silicon Lattice},
  author = {Ming, Fangfei and Johnston, Steve and Mulugeta, Daniel and Smith, Tyler S. and Vilmercati, Paolo and Lee, Geunseop and Maier, Thomas A. and Snijders, Paul C. and Weitering, Hanno H.},
  journal = {Phys. Rev. Lett.},
  volume = {119},
  issue = {26},
  pages = {266802},
  numpages = {6},
  year = {2017},
  month = {Dec},
  publisher = {American Physical Society},
  doi = {10.1103/PhysRevLett.119.266802},
  url = {https://link.aps.org/doi/10.1103/PhysRevLett.119.266802}
}

@article{Ming2023,
  title={Evidence for chiral superconductivity on a silicon surface},
  author={Ming, Fangfei and Wu, X and Chen, C and Wang, Kedong D and Mai, Peizhi and Maier, Thomas A and Strockoz, J and Venderbos, JWF and Gonz{\'a}lez, Cesar and Ortega, Jose and others},
  journal={Nature Physics},
  volume={19},
  number={4},
  pages={500--506},
  year={2023},
  publisher={Nature Publishing Group UK London}
}

@article{Ko2019,
  title={Atomic-scale manipulation and in situ characterization with scanning tunneling microscopy},
  author={Ko, Wonhee and Ma, Chuanxu and Nguyen, Giang D and Kolmer, Marek and Li, An-Ping},
  journal={Advanced Functional Materials},
  volume={29},
  number={52},
  pages={1903770},
  year={2019},
  publisher={Wiley Online Library}
}

@article{Yamaji2022,
  title={Zeros of Green functions in topological insulators},
  author={Misawa, Takahiro and Yamaji, Youhei},
  journal={Physical Review Research},
  volume={4},
  number={2},
  pages={023177},
  year={2022},
  publisher={APS}
}

@article{Hosur2025,
  title={Robust boundary Luttinger surfaces in topological band structures},
  author={Chen, Kai and Hosur, Pavan},
  journal={Physical Review B},
  volume={111},
  number={12},
  pages={125132},
  year={2025},
  publisher={APS}
}

@article{Stern2024,
  title={Ring states in topological materials},
  author={Queiroz, Raquel and Ilan, Roni and Song, Zhida and Bernevig, B Andrei and Stern, Ady},
  journal={arXiv preprint arXiv:2406.03529},
  year={2024}
}

@article{Weitering2017,
  title={Realization of a hole-doped Mott insulator on a triangular silicon lattice},
  author={Ming, Fangfei and Johnston, Steve and Mulugeta, Daniel and Smith, Tyler S and Vilmercati, Paolo and Lee, Geunseop and Maier, Thomas A and Snijders, Paul C and Weitering, Hanno H},
  journal={Physical Review Letters},
  volume={119},
  number={26},
  pages={266802},
  year={2017},
  publisher={APS}
}

@article{Weitering2023,
  title={Evidence for chiral superconductivity on a silicon surface},
  author={Ming, Fangfei and Wu, X and Chen, C and Wang, Kedong D and Mai, Peizhi and Maier, Thomas A and Strockoz, J and Venderbos, JWF and Gonz{\'a}lez, Cesar and Ortega, Jose and others},
  journal={Nature Physics},
  volume={19},
  number={4},
  pages={500--506},
  year={2023},
  publisher={Nature Publishing Group UK London}
}

@article{Qi-Zhang2011,
  title={Topological insulators and superconductors},
  author={Qi, Xiao-Liang and Zhang, Shou-Cheng},
  journal={Reviews of modern physics},
  volume={83},
  number={4},
  pages={1057--1110},
  year={2011},
  publisher={APS}
}

@article{Hasan-Kane2010,
  title={Colloquium: topological insulators},
  author={Hasan, M Zahid and Kane, Charles L},
  journal={Reviews of modern physics},
  volume={82},
  number={4},
  pages={3045--3067},
  year={2010},
  publisher={APS}
}

@article{Moore2010,
  title={The birth of topological insulators},
  author={Moore, Joel E},
  journal={Nature},
  volume={464},
  number={7286},
  pages={194--198},
  year={2010},
  publisher={Nature Publishing Group UK London}
}

@article{Vishwanath2018-RMP,
  title={Weyl and Dirac semimetals in three-dimensional solids},
  author={Armitage, N Peter and Mele, Eugene J and Vishwanath, Ashvin},
  journal={Reviews of Modern Physics},
  volume={90},
  number={1},
  pages={015001},
  year={2018},
  publisher={APS}
}

@article{Bradlyn-Cano2021,
  title={Band representations and topological quantum chemistry},
  author={Cano, Jennifer and Bradlyn, Barry},
  journal={Annual Review of Condensed Matter Physics},
  volume={12},
  number={1},
  pages={225--246},
  year={2021},
  publisher={Annual Reviews}
}

@article{Bena2025,
  title={Impurity-induced Mott ring states and Mott zeros ring states in the Hubbard operator formalism},
  author={Pangburn, Emile and Banerjee, Anurag and P{\'e}pin, Catherine and Bena, Cristina},
  journal={Physical Review B},
  volume={112},
  number={12},
  pages={125157},
  year={2025},
  publisher={APS}
}

@article{Banerjee2025,
  title={Topological charge excitations and Green's function zeros in paramagnetic Mott insulators},
  author={Pangburn, Emile and P{\'e}pin, Catherine and Banerjee, Anurag},
  journal={Physical Review B},
  volume={112},
  number={8},
  pages={085105},
  year={2025},
  publisher={APS}
}

@article{Setty2021-2,
  title={Dilute magnetic moments in an exactly solvable interacting host},
  author={Setty, Chandan},
  journal={arXiv preprint arXiv:2105.15205},
  year={2021}
}

@article{PWP2022,
  title={Discrete symmetry breaking defines the Mott quartic fixed point},
  author={Huang, Edwin W and Nave, Gabriele La and Phillips, Philip W},
  journal={Nature Physics},
  volume={18},
  number={5},
  pages={511--516},
  year={2022},
  publisher={Nature Publishing Group UK London}
}

@article{SM2026,
  title = {Supplemental Material},
  url = {https://github.com/ISU-csetty-lab/hubbard-SM/blob/main/paper_suplement.pdf}
}
 \end{document}